\documentclass[preprint]{aastex63}
\shortauthors{Vedantham et al.}
\shorttitle{Radio emission from SPI}

\begin{document}
\title{Coherent radio emission from a quiescent red dwarf indicative of star-planet interaction}
\correspondingauthor{H. K. Vedantham}
\email{vedantham@astron.nl}

\author{H. K. Vedantham}
\affiliation{ASTRON, Netherlands Institute for Radio Astronomy, Oude Hoogeveensedijk 4, Dwingeloo, 7991 PD, The Netherlands}
\affiliation{Kapteyn Astronomical Institute, University of Groningen, PO Box 72, 97200 AB, Groningen, The Netherlands}

\author{J. R. Callingham}
\affiliation{ASTRON, Netherlands Institute for Radio Astronomy, Oude Hoogeveensedijk 4, Dwingeloo, 7991 PD, The Netherlands}

\author{T. W. Shimwell}
\affiliation{ASTRON, Netherlands Institute for Radio Astronomy, Oude Hoogeveensedijk 4, Dwingeloo, 7991 PD, The Netherlands}
\affiliation{Leiden Observatory, Leiden University, PO Box 9513, 2300 RA, Leiden, The Netherlands}

\author{C. Tasse}
\affiliation{GEPI, Observatoire de Paris, Universit\'{e} PSL, CNRS, 5 place Jules Janssen, 92190 Meudon, France}

\author{B. J. S. Pope}
\affiliation{NASA Sagan Fellow, Center for Cosmology and Particle Physics, Department of Physics, New York University, 726 Broadway, New York, NY 10003, USA}

\author{M. Bedell}
\affiliation{Flatiron Institute, Simons Foundation, 162 Fifth Ave, New York, NY 10010, USA}

\author{I. Snellen}
\affiliation{Leiden Observatory, Leiden University, PO Box 9513, 2300 RA, Leiden, The Netherlands}

\author{P. Best}
\affiliation{Institute for Astronomy, Royal Observatory, Blackford Hill, Edinburgh EH9 3HJ}

\author{M. J. Hardcastle}
\affiliation{Centre for Astrophysics Research, University of Hertford-shire, College Lane, Hatfield AL10 9AB}

\author{M. Haverkorn}
\affiliation{Radboud University Nijmegen, P.O. Box 9010, 6500 GL Nijmegen, Netherlands}

\author{A. Mechev}
\affiliation{Leiden Observatory, Leiden University, PO Box 9513, 2300 RA, Leiden, The Netherlands}

\author{S. P. O'Sullivan}
\affiliation{Hamburger Sternwarte, Universit\"at Hamburg, Gojenbergsweg 112, D-21029 Hamburg, Germany}
\affiliation{School of Physical Sciences and Centre for Astrophysics \& Relativity, Dublin City University, Glasnevin, D09 W6Y4, Ireland}

\author{H. J. A. R\"ottgering}
\affiliation{Leiden Observatory, Leiden University, PO Box 9513, 2300 RA, Leiden, The Netherlands}

\author{G. J. White}
\affiliation{Department of Physics and Astronomy, The Open University, Walton Hall, Milton Keynes, MK7 6AA, UK}
\affiliation{RAL Space, STFC Rutherford Appleton Laboratory, Chilton, Didcot, Oxfordshire, OX11 0QX, UK}

\begin{abstract}
Low frequency ($\nu\lesssim 150\,$MHz) stellar radio emission is expected to originate in the outer corona at heights comparable to and larger than the stellar radius. Such emission from the Sun has been used to study coronal structure, mass ejections,  space-weather conditions around the planets \citep{schwenn-2006}. Searches for low-frequency emission from other stars have only detected a single active flare-star \citep{lynch-2017} that is not representative of the wider stellar population. Here we report the detection of low-frequency radio emission from a quiescent star, GJ\,1151--- a member of the most common stellar type (red dwarf or spectral class M) in the Galaxy. The characteristics of the emission are similar to those of planetary auroral emissions \citep{zarka-1998} (e.g. Jupiter's decametric emission), suggesting a coronal structure dominated by a global magnetosphere with low plasma density. 
Our results show that large-scale currents that power radio aurorae operate over a vast range of mass and atmospheric composition, ranging from terrestrial planets to main-sequence stars.
The Poynting flux required to produce the observed radio emission cannot be generated by GJ\,1151's slow rotation, but can originate in a sub-Alfv\'{e}nic interaction of its magnetospheric plasma with a short-period exoplanet. The emission properties are consistent with theoretical expectations  \citep{zarka-2007,lanza-2009,saur-2013,turnpenney-2018} for interaction with an Earth-size planet in a $\sim 1-5$\,day-long orbit.
\end{abstract}  
\section{main text}
\begin{figure}
    \centering
    \includegraphics[width=0.6\linewidth]{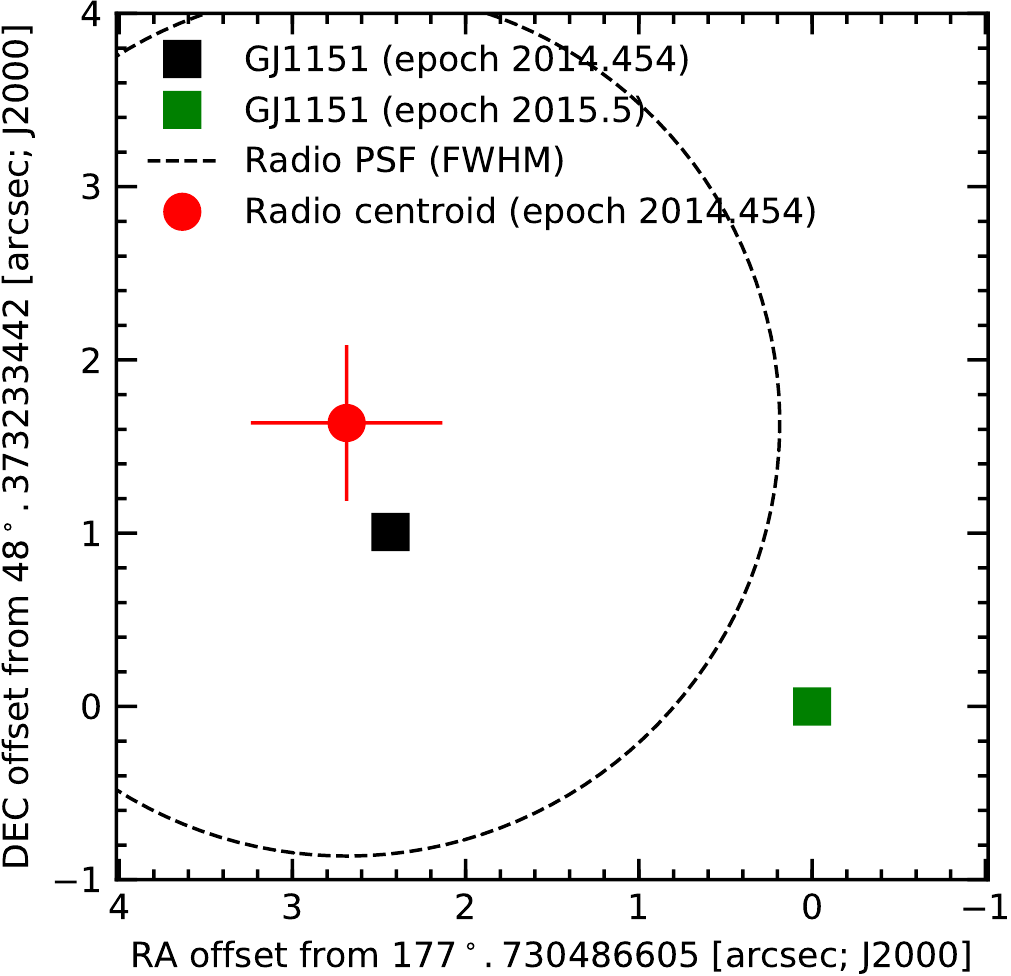}
    \caption{Relative astrometry of the radio source and the optical position of M-dwarf star GJ 1151. The optical position and proper-motion correction is based on the Gaia DR2 catalog. The error-bars show the $\pm 1\sigma$ errors on the radio source centroid that were computed by adding the formal errors in source-finding and the absolute LoTSS astrometric uncertainty in quadrature. }
    \label{fig:extfig1}
\end{figure}

We discovered radio emission in the direction of the quiescent red dwarf star GJ\,1151 by cross-matching catalogued radio sources in the LOFAR Two-Metre Sky Survey (LoTSS) data release I \citep{LoTSS-2}, with nearby stars within a distance of $d<20\,{\rm pc}$ from the {\em Gaia} DR2 database \citep{gaia-dr2}. 
The distance cut was imposed to maximise our chances of finding inherently faint stellar and planetary radio emission while maintaining a low false association rate \citep{joe_rnaas}.
We found one match at high significance: GJ\,1151 which is the closest catalogued star within the radio survey footprint.  
The radio source lies at a distance of $0''.17(55)$ in right-ascension and $0''.63(45)$ in declination from the proper motion corrected optical position of GJ1151 ($1\sigma$ errors in parentheses hereafter; see Figure \ref{fig:extfig1}). 

\begin{figure}
\begin{center}
\includegraphics[width=0.98\linewidth]{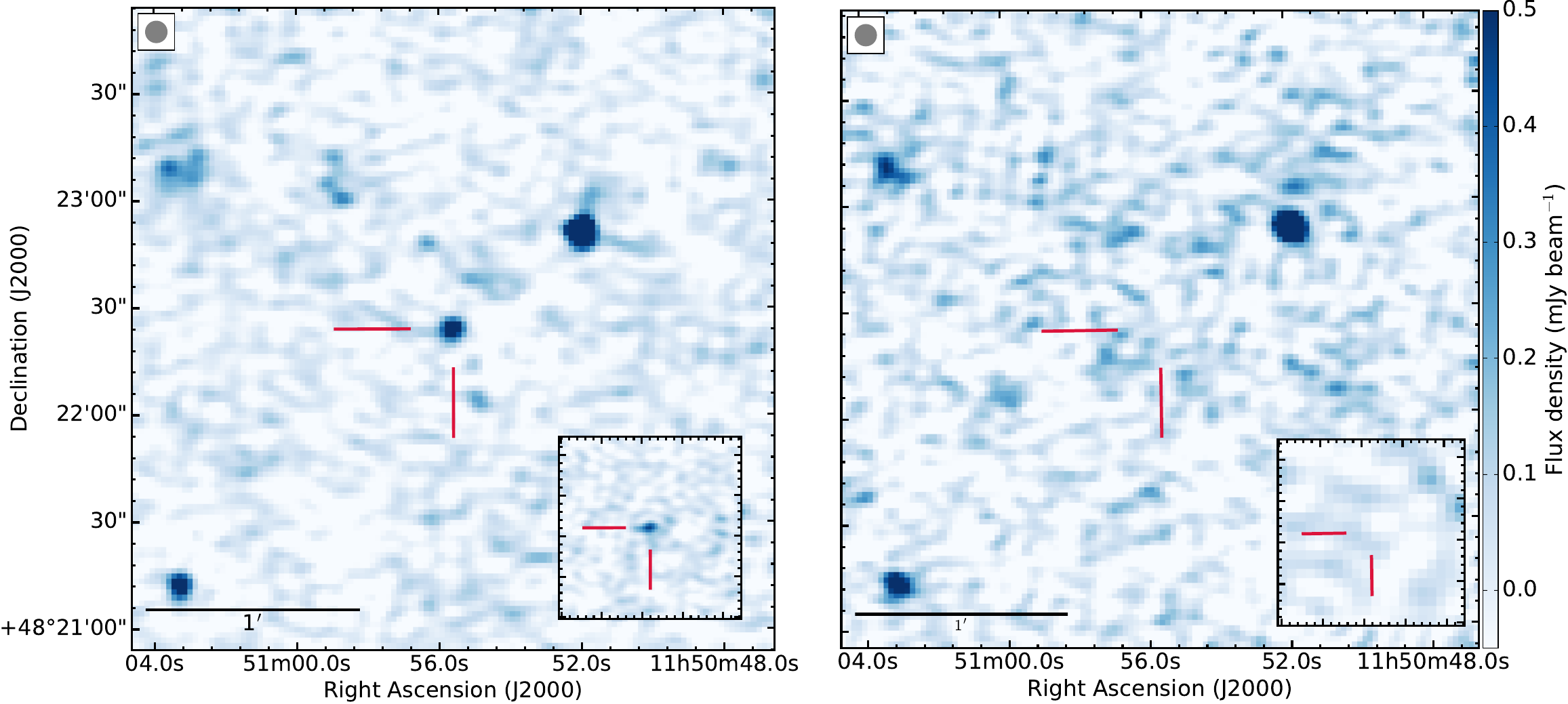}
\caption{Total intensity deconvolved images of the region around GJ\,1151 for two different epochs left panel: 2014 June 16, and right panel: 2014 May 28). The cross-hairs point to the location of GJ\,1151 (see extended Figure 1 for astrometric details). The inset in both panels displays the Stokes V (circular polarisation) image for the respective epoch. The time-frequency averaged Stokes I and V flux-densities are 0.89(8)\,mJy and 0.57(4)\,mJy  respectively. \label{fig:mainfig1}}
\end{center}
\end{figure}

GJ\,1151 was observed by four partially overlapping LoTSS pointings conducted within a span of $\sim 1$\,month. The LoTSS radio source ILT\,J115055.50+482225.2 is detected in only one, and has a high circularly polarised fraction of $64 \pm 6 \%$  (see Figure \ref{fig:mainfig1}). 
The transient nature and high polarisation fraction are inconsistent with known properties of extragalactic radio sources, but consistent with that of stellar and planetary emissions \citep{dulk-1985}. 
Based on the positional co-incidence, transient nature, and high circularly polarised fraction, we conclusively associate the radio source with GJ\,1151. The astrometric uncertainty of $\approx 0.2''$ in LoTSS data is insufficient to astrometrically differentiate between the stellar corona and a hypothetical planetary magnetosphere as the site of emission. 

\begin{figure}
\centering
\includegraphics[width=0.98\linewidth]{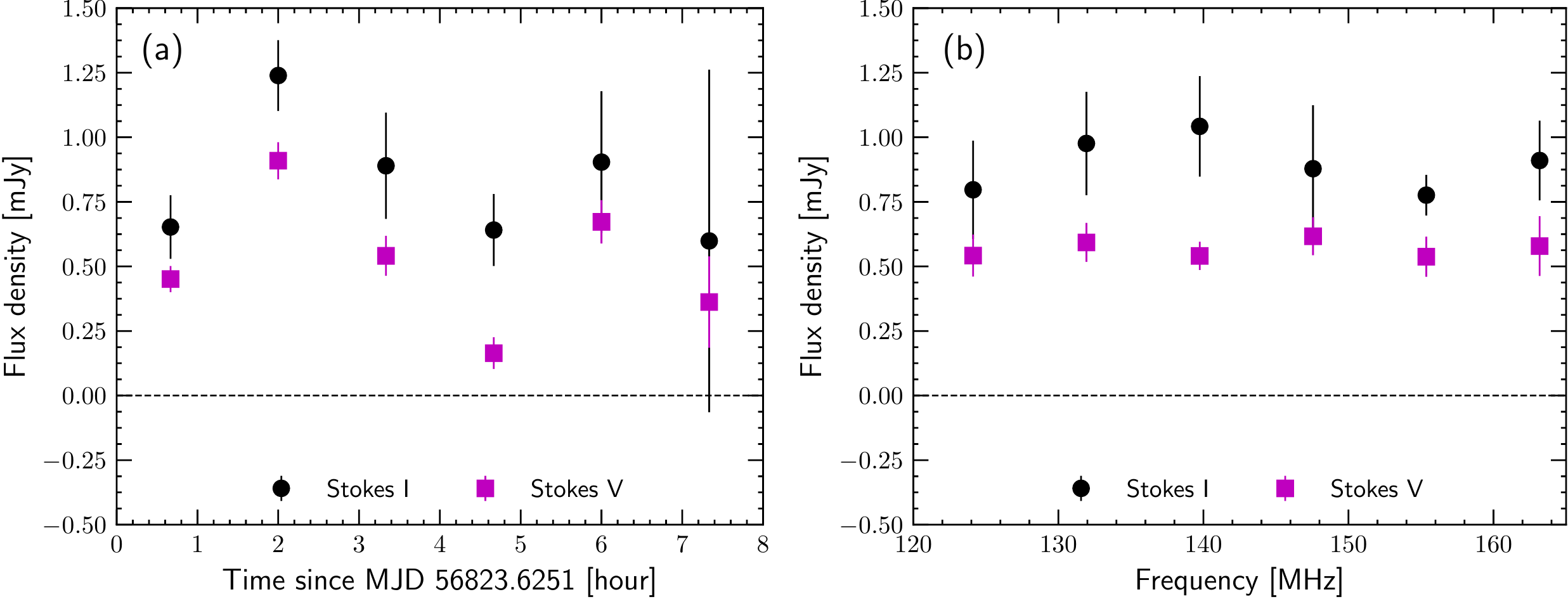}
\caption{The time and frequency variability of the total flux-density (Stokes I; black circles) and circular polarised flux-density (Stokes V; magenta squares) of the radio source in GJ\,1151. The spectrum is measured over the entire 8-hour exposure and the time-series is measured over the entire bandwidth. The error-bars span $\pm 1\sigma$.\label{fig:mainfig2}}
\end{figure}

To determine the spectro-temporal characteristics of the radio emission, we extracted its time-averaged spectrum and frequency-averaged light curve (see methods section). We found that despite temporal variability, the emission persisted for the entire $8\,{\rm hr}$ observation. 
The emission is also detected over the entire available bandwidth, $120<\nu<167\,{\rm MHz}$ ($\nu$ is the observed frequency), and has an approximately flat spectral shape (Figure \ref{fig:mainfig2}).
The in-band radio power for an isotropic emitter is $P_R\approx 2\times 10^{21}\,{\rm ergs}\,{\rm s}^{-1}$. 
The peak radiation brightness temperature is $T_b\approx 3.7\times 10^{12}x_{\ast}^{-2}\,{\rm K}$ where $x_\ast$ is the radius of the emitter in units of GJ1151's stellar radius $R_\ast\approx 1.3\times 10^{10}\,{\rm cm}$.

A unique aspect of this detected radio source is that it is associated to a star with a quiescent chromosphere. 
Stellar radio emission at gigahertz-frequencies is predominately non-thermal in origin and is powered by chromospheric magnetic activity. 
The majority of stellar radio detections are of a small class of magnetically active stars such as flare stars \citep{jackson-1989,villadsen-2018} (e.g. ADLeo), rapid rotators \citep{huges-1987} (e.g. FK Com) and close binaries \citep{umana-1998} (e.g. Algol). GJ\,1151 on the other hand is a canonical `quiescent' star, such as the Sun, based on all available chromospheric activity indicators (Table \ref{tab:comp}). 
For comparison, relatively intense broad-band noise storms on the Sun are arcmin-scale sources with brightness temperatures of $T_b\approx 10^9\,{\rm K}$  \citep{mercier-2015}. Such an emitter will be three orders of magnitude fainter than the radio source in GJ\,1151 if observed from the same distance.

In addition to the quiescent nature of GJ\,1151, the properties of the observed radio emission are distinct from prototypical stellar bursts at cm-wavelengths.
Stellar radio emission falls into two broad phenomenological categories \citep{dulk-1985}: (a) Incoherent gyrosynchrotron emission, similar to solar noise storms \citep{mercier-2015}, characterised by a low degree of polarisation, brightness temperatures of $T_b\lesssim 10^{10}\,{\rm K}$, bandwidths of $\Delta\nu/\nu\sim 1$, and a duration of many hours, and; (b) Coherent emission (plasma or cyclotron emission), similar to solar radio bursts, characterised by a high degree of circular polarisation (up to 100\%), narrow instantaneous bandwidths ($\Delta\nu/\nu\ll 1$), and a duration ranging from seconds to minutes. 
The observed emission does not fit into either of these phenomenological classes. 
It is broad-band, has a duration of $>8$ hours and highly circularly polarised. 
The closest analogue of such emission is auroral radio emission from sub-stellar objects such as planets and ultracool dwarfs \citep{zarka-1998,hallinan-2008,hallinan-2015}. 
While canonical stellar radio bursts are powered by impulsive heating of plasma trapped in compact coronal loops \citep{dulk-1985,stepanov-2001} of size much smaller than the stellar radius, radio aurorae in sub-stellar objects are driven by global current systems in a large-scale dipolar magnetic field.

To gain further insight into the nature of emission, we constrained the physical properties of the radio source from first principles. The high brightness temperature and high polarisation fraction require the emission to originate from a coherent emission mechanism. The two known classes of coherent emission in non-relativistic plasma are plasma and cyclotron emission, which lead to emission at harmonics of the plasma frequency $\nu_p$ and the cyclotron frequency $\nu_c$, respectively.

Stellar busts at cm-wavelengths have previously been successfully modelled as fundamental plasma emission from coronal loops \citep{stepanov-2001}. 
However, the emissivity of the fundamental emission drops non-linearly with decreasing frequency. For typical coronal scale heights of quiescent red-dwarfs, the height-integrated fundamental emission is restricted to brightness temperatures of $< 10^{11}\,K$ at 150\,MHz (see methods section), which cannot account for the observed emission with $T_{\rm b}\sim 10^{12}\,{\rm K}$. 
Second harmonic plasma emission has a higher emissivity at low frequencies but cannot attain the high observed level of fractional polarisation (see methods section). These inconsistencies lead us to reject plasma emission as the cause and conclude that we are observing cyclotron maser emission.

Cyclotron maser emission occurs at harmonics of the local cyclotron frequency of $\nu_c \approx 2.8 B\,{\rm MHz}$, where $B$ is the magnetic field strength in Gauss. 
It is many orders of magnitude more efficient than plasma emission \citep{wu-1979,melrose-1982}. Because the emission is inherently narrow-band, the observed broad-band emission must be the aggregate emission from regions of different magnetic field strengths within the emitter. The size of a flaring coronal loop that can accommodate such a region is comparable to or larger than the size of GJ\,1151 (see methods section). This provides additional evidence in support of global magnetospheric currents as the driver of emission as opposed to impulsively heated thermal plasma in compact coronal loops.    

Due to the high electron density in a stellar corona (as compared to planetary magnetospheres), an impediment to an auroral cyclotron maser interpretation is the gyro-resonant absorption by ambient thermal electrons at harmonics of the cyclotron frequency \citep{dulk-1985,melrose-1982}. 
Escaping radiation is obtained at coronal densities lower than $\sim 10^3\,{\rm cm}^{-3}$ and $\sim 10^6\,{\rm cm}^{-3}$ for emission at the fundamental and second harmonic, respectively (see methods section). 
These values are orders of magnitude lower than typical coronal densities of solar-type stars (F$-$ and G$-$dwarfs) and highly active flare stars \citep{stepanov-2001}. 
The coronae of X-ray dim quiescent M-dwarfs on the other hand can have significantly lower base-density, and pressure scale-heights allowing for escape-conditions to be met at heights of $1-3\,R_\ast$ where magnetospheric cyclotron maser emission is expected to originate. 
For example, adopting the empirically determined universal scaling laws for coronal parameters \citep{peres-2004}, and assuming a hydrostatic corona, we find that the escape conditions can be met in GJ\,1151 at a radius of $2R_\ast$ for coronal temperatures of $T=0.7\times 10^6\,$K and $T=1.5\times 10^6\,$K for fundamental and harmonic emission, respectively (see methods section). 
The escape criterion may also explain why analogous cm-wavelength auroral emission has previously been detected in ultracool dwarfs  \citep{hallinan-2008,hallinan-2015} but not in hotter main-sequence stars. Emission at cm-wavelengths requires a kiloGauss-level level magnetic field, which is only expected close to the stellar surface ($R\approx R_\ast$) where the high electron density  may prevent radiation escape in main-sequence stars.

Auroral cyclotron maser emission is powered by persistent acceleration of magnetically confined electrons to $\sim 10\,{\rm keV}-1\,{\rm MeV}$ energies. 
In sub-stellar objects with largely neutral atmospheres the currents are thought to be driven by two processes: (i) Breakdown of rigid co-rotation of magnetospheric plasma with the object's magnetic field either due to radial diffusion of outflowing plasma   \citep{cowley-2001}, or interaction between a rotating magnetosphere and the interstellar medium \citep{turnpenney-2017} and (iii) sub-Alfv\'{e}nic interaction of the objects magnetosphere with an orbiting body \citep{zarka-1998,saur-2013,turnpenney-2018,lanza-2009}. 
Co-rotation breakdown seen in Jupiter and ultra-cool dwarfs, which are largely observed to have rotation periods less than $\sim$3 hours, is rotation-powered and has been shown to generate a radio power of $\sim 10^{13}\,{\rm ergs}\,{\rm s}^{-1}\,{\rm Hz}^{-1}$ \citep{hallinan-2015,nichols-2012}. 
GJ\,1151 has an $\approx 3000\,$hour rotation period. Assuming coronal parameters comparable to radio-loud ultra cool dwarfs, any co-rotation breakdown in GJ\,1151 will generate a polar flux that is roughly three orders of magnitude weaker than the observed radio power of $4.3\times 10^{13}\,{\rm ergs}\,{\rm s}^{-1}\,{\rm Hz}^{-1}$. 

\begin{figure}
    \centering
    \includegraphics[width=0.98\linewidth]{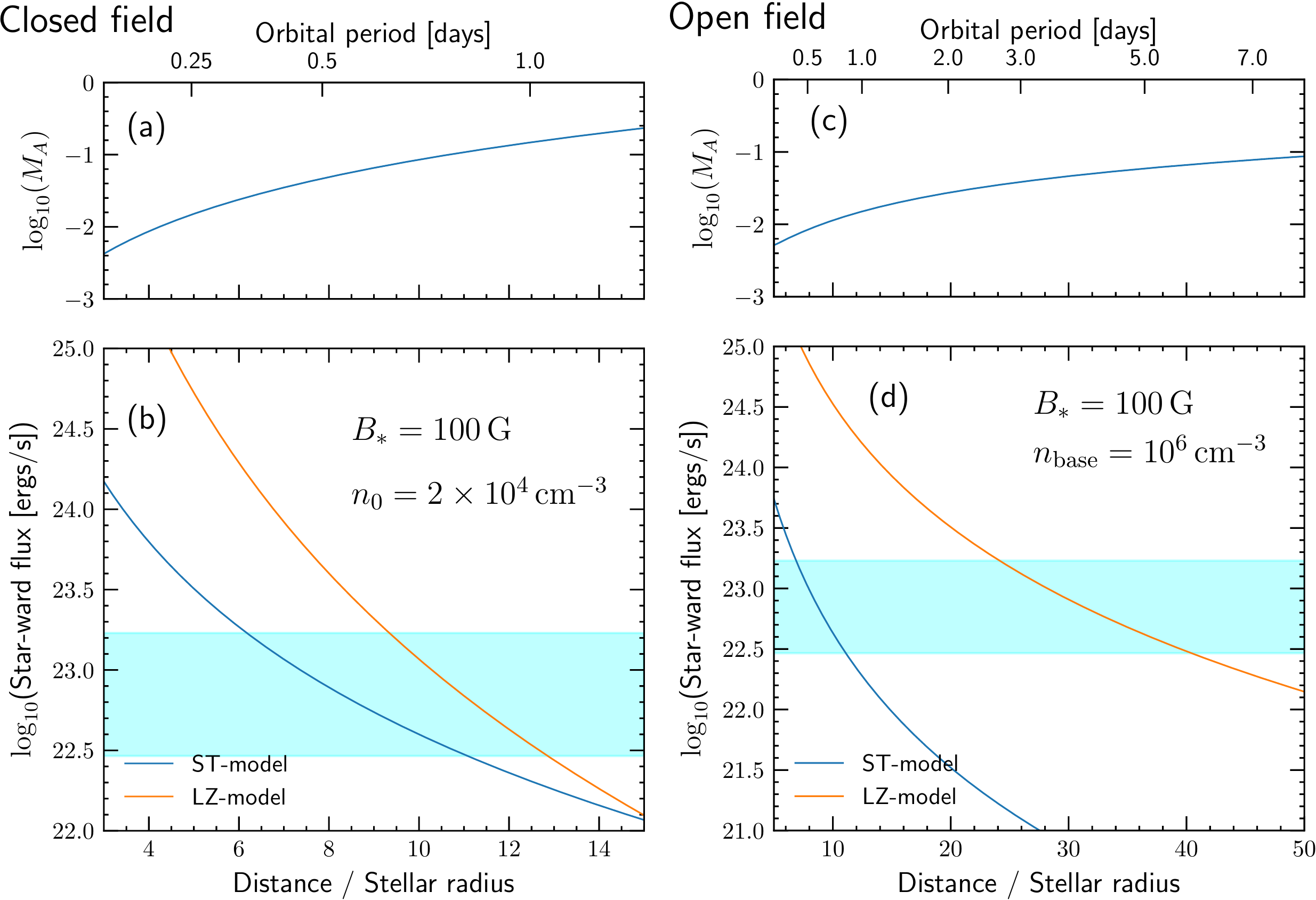}
    \caption{A comparison of observationally inferred and theoretical values for the star-ward Poynting flux from sub-Alfv\'{e}nic interaction with an Earth-size exoplanet. The left-hand panels (a and b) and right-hand panels (c and d) assume a closed dipolar and an open Parker-spiral geometries for the stellar magnetic field, respectively. The cyan rectangle is the range allowed by the observed radio flux-density. The blue and orange curves show the Poynting flux for two theoretical models of the interaction: ST-model  proposed by  \citet{saur-2013,turnpenney-2018} and LZ-model model proposed by  \citet{lanza-2009}. $B_\ast$ is the assumed surface magnetic field of the star. $n_0$ is the plasma density at the location of the putative planet (closed field case), and $n_{\rm base}$ is the base density of the coronal wind (open field case). Top panels show the Alfv\'{e}n mach number of the interaction, $M_A$. Further details are given in the methods section. \label{fig:mainfig3}}
\end{figure}

The failure of the co-rotation breakdown model points to a sub-Alfv\'{e}nic interaction as the cause of the observed radio emission. 
This scenario is a scaled-up version of the well known Jupiter-Io electrodynamic engine, and has been proposed as an avenue to study star-planet interaction \citep{zarka-2007,saur-2013,turnpenney-2018}. We checked the feasibility of this scenario by comparing theoretical estimates of the star-ward Poynting flux with that implied by the brightness of the observed emission. We considered an interaction with an Earth-like planet due to the known preponderance of such planets around red dwarf stars \citep{dressing-2015}. A planet in a one to five-day long orbit can satisfy the total energy and brightness temperature requirements for the observed radio emission (see methods section and Figure \ref{fig:mainfig3}). 

In the sub-Alfv\'{e}nic interaction scenario, although an exoplanet is implicated in the radio emission process, we have implicitly assumed that the site of emission is GJ\,1151's corona. However, a sizeable fraction of the Poynting flux intercepted by the planet can also dissipate in its magnetosphere  \citep{zarka-2007,saur-2013}. As such, the radio emission may have originated in the putative planet's magnetosphere. Recent analysis of optical signatures of star-planet interaction in short-period systems suggest that the magnetic fields of some gas-giant planets can be strong enough to generate radio emission at our observation frequency   \citep{cauley-2019}. We note however that terrestrial planets, that are more commonly found around M-dwarfs, are expected to have much weaker magnetic fields   \citep{turnpenney-2018}.

The quiescent nature of GJ\,1151 motivated us to study the phenomenology and mechanism of emission and arrive at the star-planet interaction hypothesis. Previous metre-wave observations have almost exclusively focused on highly active stars \citep{villadsen-2018, lynch-2017} making it difficult to discern possible star-planet interaction signatures with canonical stellar activity. We suggest that regardless of stellar activity level, detection of periodicity in the radio emission from GJ~1151 at a period distinct from the stellar rotation period can be used to conclusively implicate an exoplanet in the emission process with future observations. The radio-derived periodicity in such systems can additionally be corroborated against the anticipated stellar radial velocity signature. For example, our benchmark model (Earth-mass planet in a $\sim 1-5$ day orbit) implies a radial velocity signature with semi-amplitude of $\sim 1\,{\rm m/s}\times \sin i$, where $i$ is the orbital inclination of the system. Such a radial velocity signature is within the targeted sensitivity of upcoming radial velocity surveys.

We end by noting that our results show that a systematic study of the interaction between stars and short-period exoplanets using their radio emission is feasible. Based on the discovery of GJ\,1151 in a $\sim 420\,$sq. degree survey footprint, we expect many tens of such detections from the ongoing LoTSS survey, which will allow a study of star-planet interaction over different stellar types magneto-ionic interaction regimes.

\begin{table}
\caption{\label{tab:comp}The characteristics and activity indicators of GJ\,1151 compared with the prototypical radio-loud flare-star AD\,Leo. The ROSAT luminosity band is 0.1-2.4 and 0.5-8.0\,keV. All X-ray luminosities are reported for non-flare states. The upper limits reported on the X-ray luminosity of GJ\,1151 is derived from the 3$\sigma$ upper bound on the source flux. 
Distances are derived from $\emph{Gaia}$ data release 2 parallaxes \citep{gaia-dr2}. Uncertainties for literature values are reported if they are in the original text.}
\begin{tabular}{lcc}
\hline \\
{\bf Parameter} & {\bf GJ\,1151} & {\bf AD\,Leo} \\
Spectral Type & M4.5V & M3V \\
Distance (pc) & 8.04 & 4.965\\
Mass ($M_{\odot}$) & 0.17 \citep{newton-2017} & 0.42\citep{newton-2017} \\
Radius ($R_{\odot}$) & 0.2 \citep{newton-2017}  & 0.43 \citep{newton-2017}\\
H$\alpha$ equiv. width (Angs.) & $0.034\pm0.041$ \citep{newton-2017} & $-3.311\pm0.017$ \citep{newton-2017}  \\ 
H$\alpha$/Bol. lum. ($\times 10^{-4}$) & $0.067$ \citep{newton-2017} & $1.72$\citep{newton-2017} \\ 
ROSAT X-ray lum. ($\times 10^{28}$ erg s$^{-1}$) & $<0.016$  \citep{2018MNRAS.479.2351W} & $9.2 \pm 0.5$  \citep{sciortino-1999}  \\
ROSAT X-ray / Bol. lum. ($\times 10^{-5}$) & $<1.07$ \citep{2018MNRAS.479.2351W} & 105.74\citep{delfosse-1998}  \\
Rotation period (days) & $125 \pm 23$  \citep{2011ApJ...727...56I} & 2.23 \citep{2016ApJ...822...97H}  \\ 
Coronal field strength (kG) & Unknown & 0.19 \citep{2008MNRAS.390..567M}  \\ \hline \\
\end{tabular}
\end{table}

\acknowledgments 
HKV and JRC thank Prof. Don Melrose, Prof. Aline Vidotto, Prof. Phillipe Zarka for discussions. HKV thanks Dr. Vikram Ravi and Prof. Gregg Hallinan for discussions. The Leiden LOFAR team gratefully acknowledge support from the European Research Council under the European Unions Seventh Framework Programme (FP/2007-2013)/ERC Advanced Grant NEWCLUSTERS-321271. IS acknowledges funding from the European Research Council (ERC) under the European Union’s Horizon 2020 research and innovation program under grant agreement No 694513. GJW gratefully acknowledges support of an Emeritus Fellowship from The Leverhulme Trust. SPO acknowledges financial support from the Deutsche Forschungsgemeinschaft (DFG) under grant BR2026/23. MH acknowledges funding from the European Research Council (ERC) under the European Union’s Horizon 2020 research and innovation programme (grant agreement No 772663. This paper is based (in part) on data obtained with the International LOFAR Telescope (ILT). LOFAR (van Haarlem et al. 2013) is the Low Frequency Array designed and constructed by ASTRON. It has observing, data processing, and data storage facilities in several countries, that are owned by various parties (each with their own funding sources), and that are collectively operated by the ILT foundation under a joint scientific policy. The ILT resources have benefited from the following recent major funding sources: CNRS-INSU, Observatoire de Paris and Universit\'{e} d'Orl\'{e}ans, France; BMBF, MIWF-NRW, MPG, Germany; Science Foundation Ireland (SFI), Department of Business, Enterprise and Innovation (DBEI), Ireland; NWO, The Netherlands; The Science and Technology Facilities Council, UK. 
This work was in part carried out on the Dutch national e-infrastructure with the support of the SURF Cooperative through grants e-infra 160022 and 160152. The LOFAR software and dedicated reduction packages on \url{https://github.com/apmechev/GRID_LRT} were deployed on these e-infrastructure by the LOFAR e-infragroup.
This research has made use of data analysed using the University of Hertford-shire high-performance computing facility (\url{http://uhhpc.herts.ac.uk/})and the LOFAR-UK computing facility located at the University of Hertford-shire and supported by STFC [ST/P000096/1].
This work was performed in part under contract with the Jet Propulsion Laboratory (JPL) funded by NASA through the Sagan Fellowship Program executed by the NASA Exoplanet Science Institute. BJSP acknowledges being on the traditional territory of the Lenape Nations and recognises that Manhattan continues to be the home to many Algonkian peoples. We give blessings and thanks to the Lenape people and Lenape Nations in recognition that we are carrying out this work on their indigenous homelands.

\section{methods section}
\subsection{Dynamic spectrum}
To produce Figure 2, the radio data were initially processed with the standard LoTSS processing pipeline \citep{LoTSS-2} which included direction dependent instrumental gain and ionospheric corrections. The spectrum was extracted by imaging the field around GJ\,1151 using the \texttt{wsclean} software for the entire 8-hour synthesis in different six equally spaced channels. Similarly, the light curves were obtained over the entire bandwidth by splitting the 8-hour synthesis into six equal parts. The shortest baselines in the LoTSS data have larger levels of systematic errors from mis-subtracted sources. As such we conservatively chose Briggs' weighting with robustness parameter of $-0.5$ for the Stokes I images. This leads to higher noise level than naturally weighted images, but is more robust to systematic errors as it down-weights short baselines. Since the Stokes V sky is largely empty, we chose a Briggs' robustness parameter of $+0.5$ for the Stokes V images which being closer to natural weighting, yields lower noise levels.

\subsection{Plasma emission}
The free energy for plasma emission originates in electron density oscillations, called Langmuir waves, generated by a turbulent injection of impulsively heated plasma ($T_1\sim 10^8\,{\rm K}$ typically) into an ambient colder plasma ($T\sim  10^6\,{\rm K}$ typically). We used the theoretical expressions for the brightness temperature of plasma emission from \citet{stepanov-2001}. We take the Langmuir wave spectrum to be restricted to a range of wavenumbers: $k_{\rm min} = 2\pi\nu_{
\rm p}/v_1$, and $k_{\rm max} = 2\pi\nu_{\rm p}/(3v_e)$, where $v_1$ and $v_e$ are the mean velocities of the hot and cold (ambient) electrons respectively and $\nu_{\rm p}$ is the plasma frequency. For $k>k_{\rm max}$, the wave-growth is arrested by Landau damping and for $k<k_{\rm min}$ the waves cannot resonantly exchange energy with the hot electrons. 
We conservatively take the total energy density in the Langmuir waves to be $10^{-5}$ of the kinetic energy density of the ambient plasma which is the peak value obtained by both theoretical studies of non-linear effects and numerical simulations \citep{benzbook}. We assume an ambient coronal temperature of $T=2\times 10^6\,{\rm K}$ which is consistent with the X-ray non-detection of GJ1151. We assume a hydrostatic density structure with a scale height of 
\begin{equation}
    L_n \approx 6\times 10^9 (T/10^6\,{\rm K}) (R_\ast/R_\odot)^2 (M_\ast/M_\odot)^{-1},
\end{equation}
where $R_\odot$ and $M_\odot$ are the solar radius and mass respectively. We varied the hot component temperature from $5\times 10^7\,{\rm K}$ to $5\times 10^8\,{\rm K}$ and used equations 15 to 22 from  \citet{stepanov-2001} to compute the plausible range of brightness temperature for the fundamental and the harmonic. The brightness temperature of the fundamental thus calculated is between $4\times 10^9$ to $2\times 10^{11}\,$Kelvin. The brightness temperature for the harmonic is between $5\times 10^{11}$ and $1.5\times 10^{12}\,$K. Even if we assume that the entire stellar disk is filled with continuously flaring coronal loops, then the brightness temperature inferred from the observed flux density is $3.7\times 10^{12}\,$K. This alone rules out fundamental plasma emission. Even though second harmonic plasma emission can reach $\sim 10^{12}\,$K brightness temperatures, it suffers from an additional serious problem related to the high degree of polarisation observed. Solar harmonic emission has observed polarisation levels below about $20$\% \citep{benzbook}. The theory allows polarised fractions of up to $\approx 60\%$ in specific scenarios \citep{melrose-1980}.  However if coronal loops in the entire stellar disk contribute to the emission, as required by the brightness temperature constraint, then the opposing handedness of emission from regions with oppositely directed magnetic fields must lead to a substantially lower degree of net polarisation.

\subsection{Cyclotron maser from flaring coronal loop}
We consider a compact magnetic loop in the stellar corona where impulsively heated thermal plasma is injected and an unstable loss-cone distribution is set up by magnetic mirroring on either ends of the loop. For a continuously operating maser, the brightness temperature is given by \citep{melrose-1982}
\begin{equation}
T_{b} = \frac{m_ev_0^2}{4\pi k_B}\frac{\lambda^2}{Lr_0} \approx 2\times 10^{14}\left(\frac{\beta_0}{0.2}\right)^2 \left(\frac{\lambda}{200\,{\rm cm}}\right)^2\left(\frac{L}{R_\ast}\right)^{-1}\,{\rm K},
\end{equation}
where $r_0$ is the classical electron radius, $L$ is the length-scale of the trap, $m_e$ is the electron mass, $k_B$ is Boltzmann's constant, $v_0$ is the velocity of the emitting electrons and $\beta_0=v_0/c$ where $c$ is the speed of light. The emission with the above brightness temperature is centred at the ambient cyclotron frequency and is narrowband: $\Delta\nu/\nu \approx \beta^2\alpha_0^2$ where $\alpha_0$ is the opening angle of the loss-cone distribution. The observed broadband emission can be conceptually thought of as an aggregate of $\nu/\Delta\nu$ sites of emission within the magnetosphere. Consider a hypothetical magnetic trap of length $L$ and cross-sectional area of $\pi W^2 / 4$. Each site therefore has a projected area of $WL\beta^2\alpha_0^2$. Stellar coronal loops typically have $W<0.1L$ \citep{lopez-2006}. We can use these to relate the peak brightness temperature for continuous operation with the observed value to get
\begin{equation}
    L \gtrsim 16R_\ast \left(\frac{\beta}{0.2}\right)^{-4}\,\left(\frac{\alpha_0}{0.5}\right)^{-2}.
\end{equation}
Even for a high value of $\beta=0.4$ which corresponds to a plasma temperature of $\sim 10^9$\,K, we get $L\gtrsim R_\ast$. This suggests that impulsively heated thermal plasma in a compact flaring coronal loop cannot account for the observed brightness temperature.
\subsection{Escape of cyclotron maser emission}
Cyclotron maser emission must necessarily propagate through regions of decreasing magnetic field, where fundamental emission can suffer absorption at the second and higher harmonics.
Fundamental cyclotron emission is in the $x$-mode, for which the optical depth is at the $s^{\rm th}$ harmonic is \citep{melrose-1982}
\begin{equation}
\label{eqn:cmi}
\tau_s = \left(\frac{\pi}{2}\right)^{5/2}\frac{2}{c}\frac{\nu_p^2}{\nu}\frac{s^2}{s!}\left(\frac{s^2\beta^2}{2}\right)^{s-1} L_B,
\end{equation}
where $\nu_c$ is the ambient cyclotron frequency and $\beta\approx (k_BT/m_e)^{1/2} / c$ is the electron thermal velocity normalised to the speed of light. Equation \ref{eqn:cmi} must be evaluated at $\nu=s\nu_c$. The length-scale of integration is the magnetic scale height $L_B=B|\Delta B|^{-1}$ which we take to be of the order of the stellar radius $\approx 10^{10}\,{\rm cm}$. We assume a hydrostatic corona close to the star, with radial density evolution of 
\begin{equation}
    n(R) = n_{\rm b}\exp\left[-\frac{R_\ast}{L_n}\left(1-\frac{R_\ast}{R} \right) \right]
\end{equation}
where $n_{\rm b}$ is the base density and the scale height $L_n$ can be computed from the coronal temperature. 
Both of these are not observationally accessible in X-ray non detected stars such as GJ\,1151. We related the density and coronal temperature with empirically determined relationships seen in solar and stellar coronae \citep{peres-2004}: $n=4.3\times 10^6(T/10^6\,{\rm K})^{4.2}$. With this, the absorption coefficients can be computed for any coronal height once the temperature is specified.

\subsection{Radio emission from sub-Alfv\'{e}nic interaction}

\subsubsection{Energetics} 
A theoretical estimate of the star-ward Poynting flux due to the star-planet interaction is given by \citep{zarka-2007,saur-2013,lanza-2009,turnpenney-2018} (in c.g.s units)
\begin{equation}
    S^{\rm th}_{poynt} = \left[\frac{1}{2}R_{eff}^2v_{rel}B^2\right]\,\epsilon
\end{equation}
where the term in the square brackets is the incident Poynting flux on the planet, and $\epsilon$ captures efficiency factors related to the precise nature of the electrodynamic interaction (details below). $B$, $R_{eff}$, and $v_{rel}$ are, respectively, the stellar magnetic field at the location of the planet,  the effective radius of the planetary obstacle, and the relative velocity between the stellar wind flow and the planet. In convenient units, we have 

\begin{equation}
    S^{th}_{poynt} \approx 1.8\times 10^{22} \left(\frac{R_{eff}}{6000\,{\rm km}} \right)^2 \left( \frac{v_{rel}}{100\,{\rm km/s}}\right)\left( \frac{B}{1\,{\rm G}}\right)^2\left( \frac{\epsilon}{0.01}\right)\,\,{\rm ergs/s}
\end{equation}

We can compare $S^{\rm th}_{poynt}$ to the Poynting flux inferred from the observed radio emission, $S_{poynt}^{\rm obs}$, as follows. If $\Delta\nu_{tot}$ is the total bandwidth of radio emission, $\Omega$ is the beam solid angle of the radio emission, $D$ is the distance to the star and $F$ is the observed flux density, then the total emitted radio power is $P_{em} = F\Omega D^2\Delta\nu_{tot}$. We equate the total bandwidth to the peak cyclotron frequency in the star's magnetosphere: $\Delta\nu\approx 2.8B_\ast\,$MHz where $B_\ast$ is the polar surface magnetic field strength of the star. The observationally inferred star-ward Poynting flux is then $S^{obs}_{poynt}=P_{em}/\epsilon_{rad}$ where $\epsilon_{rad}$ is the efficiency with which the Poynting flux is converted to cyclotron maser emission. For the case of GJ1151, $F=0.9\,{\rm mJy}$, $D=8.04\,{\rm pc}$ which gives
\begin{equation}
    S^{obs}_{poynt} \approx 1.47\times 10^{22}\left(\frac{B_\ast}{100\,{\rm G}} \right) \left( \frac{\Omega}{0.1\,{\rm sr}}\right) \left(\frac{\epsilon_{rad}}{0.01}\right)^{-1} \,\,{\rm ergs/s}
\end{equation}

Equations 8 and 9 provide a quick check of the feasibility of the star-planet interaction model. A more detailed specification of the various free parameters is given below.

\begin{enumerate}
    \item {\em Field topology}: We consider two possible magnetic topologies at the location of the planet: a closed field geometry modelled as a dipole (planet-like), and an open field geometry that follows a Parker spiral (star-like). These correspond to the left and right panels of Figure 3 respectively. The observed emission frequency requires the surface field strength of the emitter to be $\gtrsim 60\,$G and $\gtrsim 30\,$G for emission at the fundamental and harmonic, respectively. The actual field strength of GJ\,1151 cannot be predicted accurately based on available data. We therefore assume 100\,G as a benchmark value. We note that this is broadly consistent with GJ\,1151's X-ray luminosity and rotation period (See for e.g. Figure 2 and 3 of  \citet{shulyak-2017}).

    \item {\em Nature of interaction}: For both open and closed-field cases, we consider two models to specify the interaction efficiency ($\epsilon$ in eqn. 7 and 8): (a) one of \citet{saur-2013,turnpenney-2018} called ST-model hereafter and (b) one proposed in  \citet{lanza-2009} called LZ-model hereafter. These correspond to the blue and orange lines in Figure 3 respectively. For the ST-model, $\epsilon=M_A\alpha^2\sin^2\Theta$, where $M_A$ is the Alfv\'{e}n Mach number at the planet, $\Theta$ is the angle between the stellar magnetic field at the planet and the stellar wind velocity in the frame of the planet, and $\alpha$ is the relative strength of the sub-Alfv\'{e}nic interaction. We assume that the planet has a highly conductive atmosphere for which $\alpha=1$. For the LZ-model, $\epsilon = \gamma/2$, where $0<\gamma<1$ is a geometric factor \citep{lanza-2009}. We assume the average value of $\gamma=0.5$.
 
     \item {\em  Plasma density and velocity}: For the open-field case, we assume a base coronal density of $n_{\rm base}=10^6\,{\rm cm}^{-3}$ which satisfies the radiation escape condition. Because the coronal plasma thermally expands along the open field lines, we let the base density evolve with radial distance $\propto r^{-2}$. The wind speed is assumed to follow the Parker solution with a base temperature of $10^6\,{\rm K}$. The wind speed dominates the relative velocity in the open-field case.
    
    For the closed field case, there is no substantial stellar wind at the planet's location. Due to the slow rotation of GJ\,1151, the relative velocity is largely determined by the orbital motion of the planet. We assume a constant density of $n_0=2\times 10^4\,{\rm cm}^{-3}$ at the orbital location of the planet. For comparison, the plasma density at Io's orbit is about tens times smaller and is primarily due to Io's volcanic out-gassing with negligible contribution from Jupiter itself. We have heuristically assumed a larger value as it can accommodated the presence of a tenuous stellar corona as well as an outgassing planet that is much larger than Io. In our calculation of the Alfv\'{e}n mach-number, we assume a hydrogen plasma.
    
    \item {\em Planetary parameters}: Due to the preponderance of Earth-like planets around M-dwarfs, we take the planetary radius to be 6400\,km and dipolar magnetic field with a surface strength of 1\,G. The effective radius of a magnetised planet for electrodynamic interaction, $R_{eff}$, is determined by pressure balance between the planet's magnetosphere and the stellar wind flow. Following  \citet{saur-2013}, we take this to be the radial distance from the planet at which the planetary and stellar magnetic fields are equal, times a factor of order unity that depends on the angle between the planetary magnetic-moment and the stellar field, $\theta_M$. Again following \citet{saur-2013}, we take this factor to be 1.46 and 1.73 for the open and closed field cases respectively. These correspond to $\theta_M=\pi/2$ and $\theta_M=0$ respectively. 
    \item {\em Radiation efficiency}: This factor depends on the precise nature of the electron momentum distribution which is not observationally accessible. We therefore take guidance from numerical calculations. Early calculations of the cyclotron maser instability yielded efficiencies of about 1\% \citep{aschwanden-1990}. More recent and advanced calculations yield efficiences of 10\% \citep{kuznetsov-2011} or higher. We therefore adopt a range between 1 and 10\%.
    \item {\em Beaming angle}: The total beam solid angle of the emission cone is necessary to convert emitter power to observed power. We assume an emission cone with half-opening angle $\theta$ and angular width $\Delta\theta$ which are related to the speed of the emitting electrons according to \citep{melrose-1982}: $\cos\theta\approx\Delta\theta\approx\beta$. We assume $\beta$ to line in the range $[0.3,\,0.7]$ corresponding to energies between 20\,keV and 200\,keV. The beam solid angle then lies between $0.143$ and $0.245$\,sr.
\end{enumerate}

With the above prescription the theoretically expected star-ward Poynting flux and the observationally inferred values  can be computed and contrasted as in Figure 3.

\subsubsection{Brightness temperature}
The emitting electrons powered by a star-planet interaction are largely restricted to the stellar flux tube that threads the planet. Assuming a dipolar geometry, the footprint of the flux tube on the star is an ellipse with a semi-major and semi-minor axis of
\begin{equation}
    X = \left(\frac{R_\ast}{d_p}\right)^{3/2}R_p, 
\end{equation}
and
\begin{equation}
    Y = X\left(4-\frac{3R_\ast}{d_p}\right)^{-1/2},
\end{equation}
respectively. Here $d_p$ is the orbital radius of the planet, and $R_p$ is its effective radius. The emitting region has an approximate length of order $R_\ast$ and we take the geometric mean $2\sqrt{XY}$ as its cross-sectional width. For our benchmark model of $R_p=6400\,{\rm km}$ and $d=12.5R_\ast$ the total area of the emitter normalised to the GJ\,1151's projected area is $A/(\pi R^2_\ast) \approx 0.00051$. The area of a single coherent maser site is $A\times \nu/\Delta\nu$ and the observed brightness temperature becomes $T_b = 7\times 10^{15}\,(\Delta\nu/\nu)^{-1}$. We assume a fractional bandwidth of $0.1$ corresponding to $\beta=0.5$, which leads to an intrinsic maser brightness temperature of $\sim 3\times 10^{16}\,{\rm K}$. Continuously operating masers of such high brightness temperatures can be driven by a horseshoe or shell-type electron distribution and are known to occur in magnetospheric aurorae in planets \citep{ergun-2000}.

\subsubsection{Duration and duty ratio}
Based on one detection in four exposures the duty ratio of emission is $\sim 0.25$. Because the emission lasts for $>8$\,hours, the orbital period of the planet must be larger than $\sim 1$\,day. Unlike the Jupiter--Io interaction which is seen from a special viewpoint (the Ecliptic), the range of planetary phases with visible emission is difficult to predict because it depends on (a) the inclination of the orbit, (b) magnetic obliquity, (c) and the emission cone opening angle and thickness (which in turn depend on $\beta$) which are all unknown. In addition, the source was discovered in 8-hour exposure images from a blind survey. Our detections are  therefore biased towards systems where the above factors conspire to yield a longer duration and duty-cycle of visible emission than is prototypical.

\bibliography{ref}

\end{document}